\documentclass[3p]{elsarticle}

\usepackage{latexsym}
\usepackage{epsfig}
\usepackage{amssymb}
\usepackage{color}
\newcommand{\Pg}{{\mathcal P}}
\newcommand{\Lag}{{\mathcal L}}

\newcommand{\be}{\begin{equation}}
\newcommand{\ee}{\end{equation}}
\newcommand{\lsim}   {\mathrel{\mathop{\kern 0pt \rlap
  {\raise.2ex\hbox{$<$}}}
  \lower.9ex\hbox{\kern-.190em $\sim$}}}
\newcommand{\gsim}   {\mathrel{\mathop{\kern 0pt \rlap
  {\raise.2ex\hbox{$>$}}}
  \lower.9ex\hbox{\kern-.190em $\sim$}}}

\begin{document}
\title{Cosmology with a Continuous Tower of Scalar Fields}
\author{Jose Beltr\'an Jim\'enez$^{\dag,\ddag}$}
\ead{jose.beltran@uclouvain.be}
\author{David F. Mota$^\ddag$} 
\ead{d.f.mota@astro.uio.no}
\author{ Paulo Santos$^\ddag$}
\ead{p.g.d.santos@astro.uio.no}
\address{$\;$\\
$^\dag$Centre for Cosmology, Particle Physics and Phenomenology, Institute of Mathematics and Physics, Louvain University, \\$2$ Chemin du Cyclotron 1348 Louvain-la-Neuve (Belgium)\\$\;$\\
$^\dag$D\'epartement de Physique Th\'eorique and Center for Astroparticle Physics,
Universit\'e de Gen\`eve, 24 quai Ansermet, CH--1211 Gen\`eve 4
Switzerland
\\$\;$\\
$^\ddag$Institute for Theoretical Astrophysics, University of Oslo, P.O. Box 1029 Blindern, N-0315 Oslo, Norway}

\date{\today}

\begin{abstract}
We study  the cosmological evolution for a universe in the presence of  a continuous tower of massive scalar fields which can drive the current phase of accelerated expansion of the universe and, in addition, can contribute as a dark matter component. The tower consists of a continuous set of massive scalar fields with a gaussian mass distribution. We show that, in a certain region of the parameter space, the {\it heavy} modes of the tower (those with masses much larger than the Hubble expansion rate) dominate at early times and make the tower behave like the usual single scalar field whose coherent oscillations around the minimum of the potential give a matter-like contribution. On the other hand, at late times, the {\it light} modes (those with masses much smaller than the Hubble expansion rate) overcome the energy density of the tower and they behave like a perfect fluid with equation of state ranging from 0 to -1, depending on the spectral index of the initial spectrum. This is a distinctive feature of the tower with respect to the case of quintessence fields, since a massive scalar field can only give acceleration with equation of state close to -1. Such unique property is the result of a synergy effect between the different mass modes. Interestingly, we find that, for some choices of the spectral index, the tower tracks the matter component at high redshifts (or it can even play the role of the dark matter) and eventually becomes the dominant component of the universe and give rise to an accelerated expansion.
\end{abstract}


\maketitle

\section{Introduction}

More than a decade after the first evidences for the accelerated expansion of the universe \cite{SN98}, the true underlying mechanism responsible for it still remains unclear. After all this time, a simple cosmological constant still seems to succeed in explaining most of the cosmological observations by using one single additional parameter \cite{Komatsu}. This has lead to establish the so-called $\Lambda$CDM model as the standard model of cosmology, in which we include the cosmological constant $\Lambda$ accounting for about 70\% of the energy content of the universe and the Cold Dark Matter component comprising 25\% of the universe content, that amounts to $85\%$ of the matter component. Despite its remarkable phenomenological success, alternatives to the standard $\Lambda$CDM model have been extensively explored because it is not completely satisfactory from a pure theoretical point of view. In particular,  we need to assume that 95\% of the cosmological budget is due to unknown fields or exotic forms of energy. Concerning the dark energy sector, alternatives to explain the accelerated expansion have been considered in part due to the  tiny required value for $\Lambda$, which turns out to be many orders of magnitude smaller than its expected {\it natural} scale (see \cite{Martin:2012bt} for an extensive review on the cosmological constant problem). This naturalness problem is also linked to the {\it coincidence} problem referred to the fact that dark matter and dark energy give comparable contributions to the total energy density of the universe precisely today, while they have evolved very differently throughout the universe expansion history.

A simple step forward is to assume that the accelerated expansion is not caused by a cosmological constant, but it is driven by some dynamical field \cite{copeland} whose evolution could solve the naturalness and/or coincidence problems of the cosmological constant. Of course, in these alternative set-ups, the cosmological constant is assumed to be set to zero by some unknown mechanism. The most widely studied alternative models for dark energy are by far those based on scalar fields with a certain potential (quintessence) or non-standard kinetic terms (K-essence) \cite{K-essence}, very much like in the inflationary models for the early universe. However, also higher spin fields have been considered as dark energy candidates, like vector fields \cite{vectorDE} or general $p$-forms \cite{3formsDE}. The difference between inflationary models and dark energy models are the involved scales and that, whereas during inflation the accelerated expansion must end, for dark energy models is desirable to have the accelerated solutions as attractors so that the natural evolution of the universe is towards an accelerated phase. Besides its simplicity, scalar fields are a  common feature of high energy physics that extend the standard model of particles,  so it is natural to think that they can be present in the universe and play a relevant role in the cosmological evolution.  An alternative explanation for the cosmic acceleration could also be that the gravitational sector is modified on large scales, i.e., the validity of General Relativity could breakdown in the infrared regime. In this context we could mention the so-called $f(R)$ theories \cite{f(R)} or those based on extra dimensions like the DGP model \cite{DGP}. It is however a typical feature of modified theories of gravity that the new effects can be captured by a certain scalar degree of freedom like $\partial f(R)/\partial R$ in the $f(R)$ theories or the bending mode in braneworld models. This is so because any modification of General Relativity necessary leads to the appearance of new degrees of freedom and this generally results in a scalar field.

Although the kind of models mentioned in the previous paragraph can give equally consistent explanations for the cosmic acceleration as a cosmological constant does, naturalness problems usually still remain. For instance, in the simplest case of quintessence with a canonical massive scalar field, we need its mass to be $m\lsim H_0\simeq10^{-33}$ eV,  again a tiny value when compared to what one would expect from high energy physics \cite{copeland}. Moreover, such a value is unstable when including quantum corrections. This constraint on the effective mass is necessary  for the potential energy to dominate over the kinetic energy of the field today and thus its equation of state is near $-1$. The situation does not improve much by resorting to more sophisticated potentials or non-standard kinetic terms. Most of the times, the naturalness problems reappear hidden behind some parameters of the potential.

On the other hand, dark matter models resorting to scalar fields also exist. For instance, the lightest massive mode of a Kaluza-Klein (KK) tower corresponding to the compactification of one extra dimension is a viable candidate for dark matter. Stability of such a particle is guaranteed by KK-parity corresponding to momentum conservation associated to spatial translation in the fifth dimension. Also resorting to extra dimensions we can mention the vibrations of our brane in the extra dimension (branons) that can play the role of dark matter \cite{branons}.  More recently, an ensemble of scalar particles which can interact among them has been considered as a dark matter candidate \cite{TDM}. The interesting feature of this approach is that the scalar particles do not need to be stable, but a certain balance between the width of the decays and the abundances allows to have a dynamical scenario in which the different effects conspire to give an overall dark matter contribution.  In \cite{Chialva:2012rq}, it was also considered a scenario with multiple Kaluza-Klein dark matter candidates.

In this work we present a novel approach to the dark energy problem based on scalar fields that, in addition, allows to unify it with the dark matter component. Unlike previous studies, we do not rely on one single scalar field, nor a finite discrete collection of (perhaps interacting) scalar fields. Here we shall study a cosmological scenario with a continuous tower of massive scalar fields which only interact with each other by means of gravity. Towers of scalar fields arise in a natural manner in nonlocal theories \cite{nonlocal,nonlocal2,nonlocal3} or theories with extra dimensions as we have discussed above. If one considers one compact extra dimension, a massless scalar field leads to a spectrum in the 4 large non-compact dimensions consisting of a massless scalar field plus a discrete tower of massive scalar fields, being the corresponding masses determined by the compactification scale of the extra dimension. Moreover, the mass gap between two consecutive modes of the tower is inversely proportional to such a scale. Thus, as the size of the extra dimension gets larger, the mass gap between the different components of the tower gets smaller and, in the limit when the extra dimension is infinite, the tower becomes continuous. This actually happens in models with extra large dimensions like the Randall-Sundrum or DGP models. Another context in which a continuous distribution of massive fields appear is as an effective description of the unparticle fields \cite{unparticles1}. 

Here however we shall adopt a more phenomenological approach and shall simply assume the presence of a continuous tower of massive scalar fields in our universe, without paying special attention to the underlying mechanism generating it, that could be either unparticle physics or braneworld scenarios, or even more exotic set-ups. Our aim is to show that a continuous tower of simple massive scalar fields can lead to accelerated expansion and, in addition, it can simultaneously give a dark matter contribution and, thus, allowing for the unification of both dark components of the universe. It is important to keep in mind that the specific generating procedure for the tower could also give an additional contribution to the cosmological constant (as it happens for instance with KK towers). We do not attempt to tackle the cosmological constant problem here and we shall assume that such contributions do not play a relevant role, as it is done in most dark energy models in the literature.

The Letter is organised as follows. In the next section we give our starting action for the continuous tower that we shall use for our analysis and we will derive the relevant equations. In Section 3 we study the cosmological evolution of the tower during the standard radiation and matter dominated epochs and show how the light modes of the tower eventually dominate the universe and can produce an accelerated expansion. In Section 4 we study the case in which the tower dominates the universe and show how we can indeed have a transition from matter domination to accelerated expansion in a tower dominated universe. Finally, in Section 5 we discuss the results of the Letter.


\section{A continuous tower}
As we have mentioned in the Introduction, we are not concerned about the true underlying mechanism generating the tower, but we shall adopt a phenomenological approach. Thus, we simply assume that our tower is described by the action
\be
\Lag_\phi=\frac{1}{2}\int dm\Pg(m)\left(\partial_\mu\phi_m\partial^\mu\phi_m-m^2\phi_m^2\right)
\ee
where $\Pg(m)$ is the spectral mass distribution of the tower, that we assume to be gaussian with zero mean
\be
\Pg(m)=\frac{1}{\sqrt{2\pi}\sigma}e^{-\frac{m^2}{2\sigma^2}}.
\ee
It is interesting to note that the only free parameter present in the action is the dispersion of the distribution $\sigma$. Notice that, although $m$ runs over the real numbers, no tachyonic modes are actually present because $m^2$ is always positive. In fact, since the mass distribution peaks at $m=0$, the corresponding equations will remain invariant under the change $m\rightarrow -m$, i.e., we have a {\it mass reversal symmetry} and we could consider just $m\geq0$ without affecting the physics of the system. One could of course consider different spectral mass distributions like a gaussian distribution for $m^2$, a $\chi^2$-distribution or even more sophisticated distributions. However, the key point for our purposes here is that the distribution peaks at the origin\footnote{This also prevents the introduction of additional dimensionfull parameters in the theory to specify where the distribution peaks. In our case, one could use $\Pg(m)\propto {\rm exp}\Big[-\frac{(m-m_0)^2}{2\sigma^2}\Big]$ that depends on the mass $m_0$ giving the position of the maximum of the distribution. Even though we do not consider this possibility here (in order not to introduce unnecessary complications in our analysis), as long as $m_0\lsim\sigma$ our findings will still be valid. } and has a given width, being the particular details of the precise distribution under consideration not really relevant. The reason for that is that we want the lightest modes of the tower to play a relevant role in the cosmological evolution so that they can eventually drive an accelerated expansion, as we show below. Actually, having a distribution such that the origin is within the 1-$\sigma$ region would suffice for the very light modes not to be irrelevant. 


On the other hand, we are also considering the simplest case in which the different modes of the tower are decoupled from each other. In the most general set-up, the different mass modes could be coupled, although one can redefine the fields to use {\it normal} coordinates so that we end up with the above {\it diagonal} action. More explicitly, we could start from the action
\be
S=\frac{1}{2}\int d^4x\sqrt{-g}\int dm\int dn \Big[K(m,n)\partial_\mu\psi_m\partial^\mu\psi_n-V(m,n)\psi_m\psi_n\Big].
\ee
Then, we can redefine the fields as follows:
\be
\psi_m=\int A(m,p)\phi_p dp
\ee
with $A(m,p)$ satisfying\footnote{Assuming that both operators $K(m,n)$ and $V(m,n)$ are symmetric it is always possible to diagonalise the action.}
\begin{eqnarray}
\int dp dq A(m,p)A(n,q) K(p,q)&=&\Pg(m)\delta(m-n),\nonumber\\
\int dp dq A(m,p)A(n,q) V(p,q)&=&m^2\Pg(m)\delta(m-n)
\end{eqnarray}
with $\delta(m-n)$ the delta function. Then, it is straightforward to see that this redefinition diagonalises our action so that we end up with
\be
S=\frac{1}{2}\int d^4x\sqrt{-g}\int dm\Pg(m)\left[\partial_\mu\phi_m\partial^\mu\phi_m-m^2\phi_m\phi_m\right]
\ee
which is our initial action. Notice that this is, in principle,  doable only for the quadratic part of the action. However,  if we had interaction terms, the different modes will be unavoidable coupled. On the other hand, one could always redefine the fields to use canonically normalised fields $\phi\rightarrow\phi/\sqrt{\Pg(m)}$ so that we get rid of the distribution $\Pg$ from the action. Then, the effect of the mass distribution can be transferred to the initial conditions. However, we prefer to keep $\Pg(m)$ in the action so that we can differentiate its effect form that of the initial conditions.

In the present case, each mode satisfies a decoupled Klein-Gordon equation
\be
(\Box +m^2)\phi_m=0.
\ee
In flat spacetime, each component of the tower evolves like a free field and they are not affected by each other. The same happens in a curved background when they are treated as test fields, i.e., we ignore their backreaction on the spacetime. However, when their gravitational contribution is relevant, they become coupled by means of gravity through Einstein equations.

The energy-momentum tensor of the tower is simply given by the superposition of the energy-momentum tensor of its components weighted by the mass distribution
\be
T_{\mu\nu}^\phi=\int dm \Pg(m)\Big[\partial_\mu\phi_m\partial_\nu\phi_m-g_{\mu\nu}\left(\partial_\mu\phi_m\partial^\mu\phi_m-m^2\phi_m^2\right)\Big].
\ee
Thus, the energy density and pressure are
\begin{eqnarray}
\rho_\phi&=&\int dm \Pg(m)\left(\partial_\mu\phi_m\partial^\mu\phi^m+m^2\phi_m^2\right)\\
p_\phi&=&\int dm \Pg(m)\left(\partial_\mu\phi_m\partial^\mu\phi^m-m^2\phi_m^2\right)
\end{eqnarray}
 so that the equation of state reads
 \be
 w_\phi\equiv\frac{p_\phi}{\rho_\phi}=\frac{\int dm \Pg(m)\left(\partial_\mu\phi_m\partial^\mu\phi^m-m^2\phi_m^2\right)}{\int dm \Pg(m)\left(\partial_\mu\phi_m\partial^\mu\phi^m+m^2\phi_m^2\right)}.
\ee
Now that we have all the equations that we need, we proceed to study the cosmological evolution of the tower in the next section.

\section{Cosmological evolution}
In a Friedmann-Lema\^itre-Robertson-Walker (FLRW) universe described by the line element  $ds^2=dt^2-a(t)^2d\vec{x}^2$, with $a(t)$ the scale factor, the homogeneous mode of each component of the tower evolves according to the equation
\be
\ddot{\phi}_m+3H\dot{\phi}_m+m^2\phi_m=0
\ee
where $H(t)=\dot{a}/{a}$ is the Hubble expansion rate. Here we see that the different modes only couple to each other when the tower contributes in a non-negligible way to the Hubble expansion rate. Since, in this section, we shall assume that the universe expansion is driven by some background barotropic fluid (radiation or matter), the different modes will evolve independently of each other.

The energy density in this case reads
\be
\rho_\phi=\frac12\int dm \;\Pg(m)\Big(\dot{\phi}_m^2+m^2\phi_m^2\Big).
\ee
If we now consider a power law expansion with $H=p/t$ (as it happens during the radiation and matter eras with $p=1/2$ and $p=2/3$ respectively), we find the usual result
\be
\phi_m(t)=t^{-\nu}\left[A_m J_\nu(|mt|)+B_mY_\nu(|mt|)\right]
\ee
with $\nu=(3p-1)/2$, $J_\nu$ and $Y_\nu$ the Bessel functions of first and second kind, and $A_m$ and $B_m$ integration constants that will, in general, depend on the mass of the corresponding mode. In particular, for the radiation era ($p=1/2$) we have 
\be
\phi^R_m(t)=t^{-1/4}\left[A_m^R J_{1/4}(|mt|)+B_m^RY_{1/4}(|mt|)\right],
\label{Rsol}
\ee
whereas for the matter era ($p=2/3$) we find
\be
\phi^M_m(t)=t^{-1}\left[A^M_m \cos(|mt|)+B^M_m\sin(|mt|)\right].
\label{Msol}
\ee
The different  masses of the modes will make them evolve in different ways so that they will contribute differently to the tower energy density. Depending on the ratio between the mass and the Hubble expansion rate, we have two regimes. For $m\ll H$ ($mt\ll1$), $\phi_m$ is approximately constant, whereas for $H\ll m$ ($mt\gg1$) the field will oscillate with  decaying amplitude. Thus, since the value of $H$ decreases as the universe expands, the natural behaviour of a given component of the tower is to remain frozen (or, more precisely, in slow roll phase) at early times until  $H\sim m$, when the field starts oscillating around the minimum of the quadratic potential. 

In the regime in which the field remains frozen, the energy density of the corresponding mode is constant, while in the oscillating regime the field behaves like a matter fluid with an equation of state that averages to zero over one oscillation, i.e., $\bar{w}=0$. This is the well-known result in the usual quintessence models. However, having a continuous tower of scalar fields, this is not the end of the story, since the total energy density (or the effective equation of state) comprises the contribution from fields with {\it any} mass. In fact, this is a crucial difference because the problem for one single scalar field to play the role of dark energy is that its mass should be {\it unnaturally} small ($\sim 10^{-33}$ eV). For the case of the continuous tower, we have {\it all} the possible masses and, in particular, those tiny masses are always there. This is true for our distribution choice, because for distributions that peak at some finite mass $m_0$ and such that the origin is far away from the maximum, the light modes will be very suppressed and their role will be irrelevant. This is the reason why we need the peak not to be many sigmas away from the origin.

\begin{figure*}
\includegraphics[width=8cm]{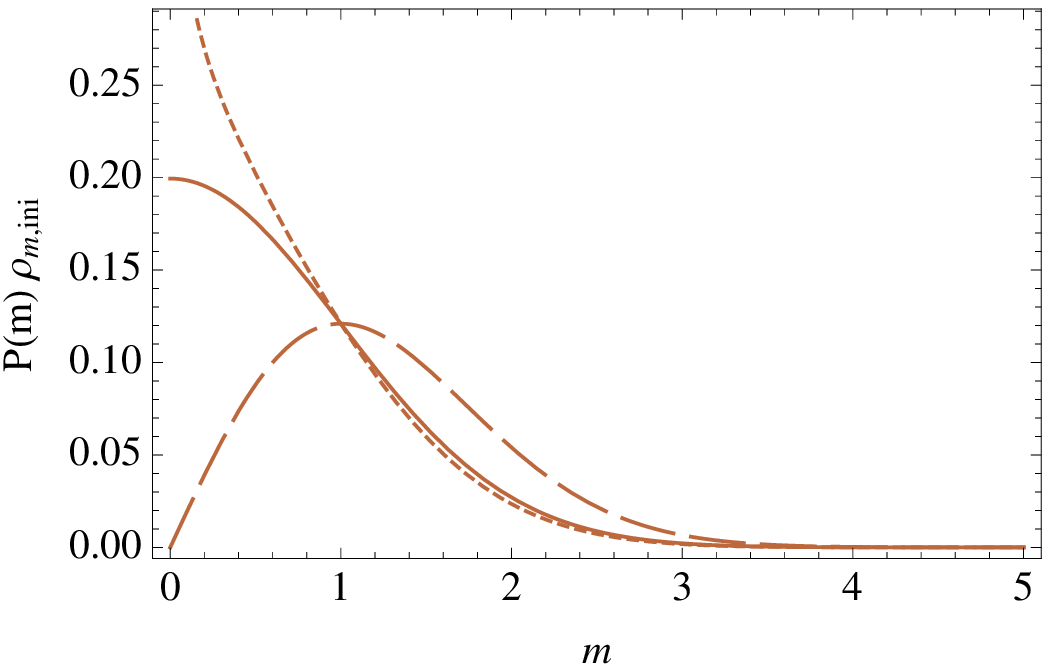}\hspace{0.4cm}
\includegraphics[width=8cm]{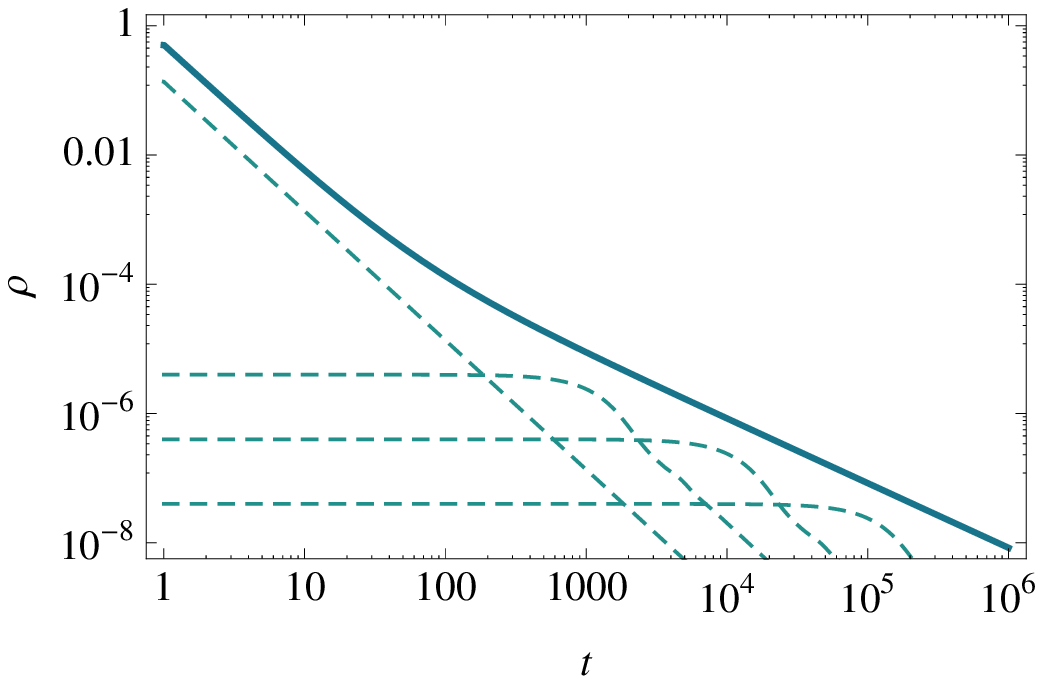}
\caption{In the left panel we show the initial energy density spectral distribution, i.e., $\Pg(m)\rho_m(t_{ini})$ for $b= 0.5,1$ and $1.2$ (long-dashed, solid and short-dashed respectively). There are two qualitatively distinctive initial distributions depending on $b$. If $b<1$, the distribution peaks at some non-vanishing value, whereas for $b\geq1$, the distribution peaks at the origin. In the right panel we show the evolution of the tower energy density in a matter dominated universe. We have chosen
 $b=1$ and $\sigma=10 H_{ini}$ so that the heavy modes that behave like a matter fluid dominate initially. We also show the evolution of {\it individual} components of the tower (dashed lines) to illustrate why the modes that are initially frozen and start decaying like a matter fluid at some time lead to a slower decay than the one corresponding to a single massive field. As we can see, the mechanism for this to happen is that we have a continuous tower with all the possible masses so that there will always be a lighter field that will decay later, making the overall tower energy density to have an equation of state different from 0. }
\label{fig1}
\end{figure*}

According to the discussion of the previous paragraph, we can decompose the tower in two sets of modes, those with masses smaller than the expansion rate (that will evolve as a cosmological constant)  and those with masses larger than the expansion rate (which will contribute as a matter fluid). However, the overall behaviour of the tower will not be as simple as this picture might suggest, since we have additional effects. To see these effects more clearly, we will split the energy density of the tower as the contribution of the {\it light} modes plus the contribution of the  {\it heavy} modes, where by {\it light} and {\it heavy} we mean with respect to the expansion rate. In other words, we can write $\rho=\rho_l+\rho_h$ with
\begin{eqnarray}
\rho_l=\int_{|m|\lsim H}dm\Pg(m)\rho_m(t)\qquad{\rm and}\qquad
\rho_h=\int_{|m|\gsim H}dm\Pg(m)\rho_m(t).
\end{eqnarray}
First of all, only those modes with masses $m\lsim \sigma$ will contribute in a non-negligible way to the tower because of the exponential suppression due to $\Pg(m)$. This means that $\rho_{h}$ is determined by the modes with $H\lsim m \lsim \sigma$. In particular, if $H>\sigma$, $\rho_h$ will be exponentially suppressed. However, since $H$ decreases throughout the universe evolution, $\rho_h$ will acquire contributions that are not suppressed by $\Pg(m)$ after the Hubble expansion rate crosses the value of the tower dispersion $\sigma$. Obviously, the cosmological evolution of the full tower depends on how the different mass modes contribute to the full energy density of the tower at each time and this will depend on the initial distribution. In order to set the initial conditions for the tower we should know the underlying mechanism leading to the generation of the tower so that we can compute the actual primordial spectrum of the tower. However, since we are not concerned here with the precise mechanism generating the tower, we shall assume a general power law primordial spectrum so that the initial conditions will be assumed to be given by
\be
\phi_m(t_{ini})=A\left\vert\frac{\sigma}{m}\right\vert^b\label{IC}
\ee
with $A$ some amplitude and $b$ the spectral index of the primordial tower distribution.  Moreover, the field will be assumed to be initially at rest so that $\dot{\phi}_m(t_{ini})=0$. On the other hand, since we want both the light and the heavy modes to contribute in a non-negligible way at the initial time, we shall further assume that $\sigma\gsim H_{ini}\sim t_{ini}^{-1}$. Finally, we should note that, in order to avoid divergences in the energy density of the tower we must impose $b\leq 3/2$. To understand why this is so, we compute the initial energy density
\be
\rho_{ini}=\frac 12\int dm\; m^2\phi^2_m(t_{ini})\Pg(m)=\frac{\sigma^2 A^2}{2\sqrt{2\pi}}\int dy\frac{e^{-y^2/2}}{y^{2(b-1)}}
\label{rhoini}
\ee
where $y\equiv\frac{m}{\sigma}$. The integral will not have convergence problems for $y\rightarrow\infty$ thanks to the exponential suppression due to $\Pg(m)$. However, we can find divergences near the origin because the integrand behaves like $y^{-2(b-1)}$ that diverges for $2(b-1)\geq1$. Precisely for this reason, we shall restrict to the case with $b < 3/2$. In Fig.\ref{fig1}, we show the initial contribution of the different modes to the tower energy density for different values of $b$.

It is clear that $\rho_h$ will always give a contribution decaying like a matter fluid, i.e., $\rho_h\propto a^{-3}$ because, for our initial conditions which assume that $\sigma\gg H_{ini}$, the dominant contribution to $\rho_h$ comes from the modes with $m\simeq\sigma$. In fact, if we approximate the gaussian by a step function we have
\be
\rho_{ini}\simeq\frac{A^2\sigma^2}{\sqrt{2\pi}}\int_0^{1} \frac{dy}{y^{2(b-1)}}
=\frac{A^2\sigma^2}{\sqrt{2\pi}(3-2b)}.
\label{rhoini2}
\ee
Then,  the fact that $H$ varies (and more modes contribute to $\rho_h$) will not affect its cosmological evolution. Moreover, since we want the heavy modes to dominate over the light ones at the initial time (we shall make this statement more precisely below), we can write
\be
 \rho_h\simeq\rho_{ini}\left(\frac{a_{ini}}{a}\right)^{3}\simeq\frac{A^2\sigma^2}{\sqrt{2\pi}(3-2b)}\left(\frac{a_{ini}}{a}\right)^{3}.
 \label{rhoh} 
 \ee
 In this expression we see again why we must restrict to cases with $b<3/2$. Otherwise, we would find a divergence.
 
For the light modes however  the variation of the amount of modes contributing to $\rho_l$ will impact its cosmological evolution because the corresponding integral will also be dominated by the upper limit which now is given by $H(t)$, a time-dependent quantity. Physically, it is easy to understand the mechanism governing the evolution of the light modes contribution. We show it explicitly in the right panel of Fig.\ref{fig1}, where we plot the evolution of the entire tower in a matter dominated universe and also show the evolution of {\it single} modes. We see how the modes that are initially frozen when their masses are larger than $H$, give a matter-like contribution when $H$ becomes smaller than the corresponding mass. However, since we have a continuous of mass modes, we have modes that start decaying at {\it any} given time. In other words, if a mode of mass $m$ starts decaying at time $t_m$, then there is a mode with mass $m+\Delta m$ that starts decaying at $t_m+\Delta t_m$ so that the final contribution to the energy density is the envelope of all the individual energy densities, as shown in the figure. We can also estimate this analytically. To compute the evolution of the light modes that can give a dark energy contribution, we write
\begin{eqnarray}
\rho_l
\simeq\frac 12\int_{|m|\lsim H} dm\; m^2\phi^2_m(t_{ini})\Pg(m)
\simeq\frac{\sigma^2 A^2}{\sqrt{2\pi}(3-2b)}\left(\frac{H}{\sigma}\right)^{3-2b}=\rho_{ini}\left(\frac{H}{\sigma}\right)^{3-2b}
\label{rhol1}
\end{eqnarray}
where we have used that the light modes remain approximately constant and their values are given by (\ref{IC}). We have also assumed that $H\ll\sigma$ so that the exponential can be approximated by unity in the integrand.  From the above expression, we see that the necessary condition for the heavy modes to dominate at the initial time $\rho_l(t_{ini})<\rho_h(t_{ini})$ is simply $\sigma>H_{ini}$ (for $3-2b>0$) because this will guarantee that $\rho_l\leq\rho_{ini}\simeq\rho_h(t_{ini})$.  Now, if we use that at high redshifts we have $H\simeq H_0\sqrt{\Omega_r}(1+z_{ini})^2$, with $\Omega_r$ the density parameter of radiation, we find that the following bound must be fulfilled 
\be
\sigma\gg H_0\sqrt{\Omega_r}(1+z_{ini})^2\simeq 3\times10^{-36}(1+z_{ini})^2\;{\rm eV}.
\ee

On the other hand, from (\ref{rhol1}) we easily obtain the effective equation of state parameter for the light modes. We see that they give a contribution to the energy density evolving as
\be
 \rho_l\propto t^{2b-3} \propto a^{\frac{2b-3}{p}}
 \label{rhol}
 \ee
where we have used that $a\propto t^p$. Then, the effective equation of state of the light modes contribution can be straightforwardly computed and we obtain
\be
w_l=-\frac{2}{3p}b+\frac1p-1
\label{wlp}
\ee
 It is interesting to notice that, irrespectively of the background expansion determined by $p$, we obtain that for $b\rightarrow3/2$ (which is our limiting case), the equation of state parameter goes to $w_l\rightarrow-1$. Of course, the result that we have obtained here is only valid for values of $b$ such that $w_l\leq0$, i.e.,  for $b\geq\frac32(1-p)$ (assuming positive $p$) because for smaller values of $b$ the light modes will also contribute as a matter component and its energy density cannot decay slower than that. In particular, for radiation ($p=1/2$) we obtain $b>3/4$, whereas for matter ($p=2/3$) the condition reads $b>1/2$.

\begin{figure}
\includegraphics[width=8.cm]{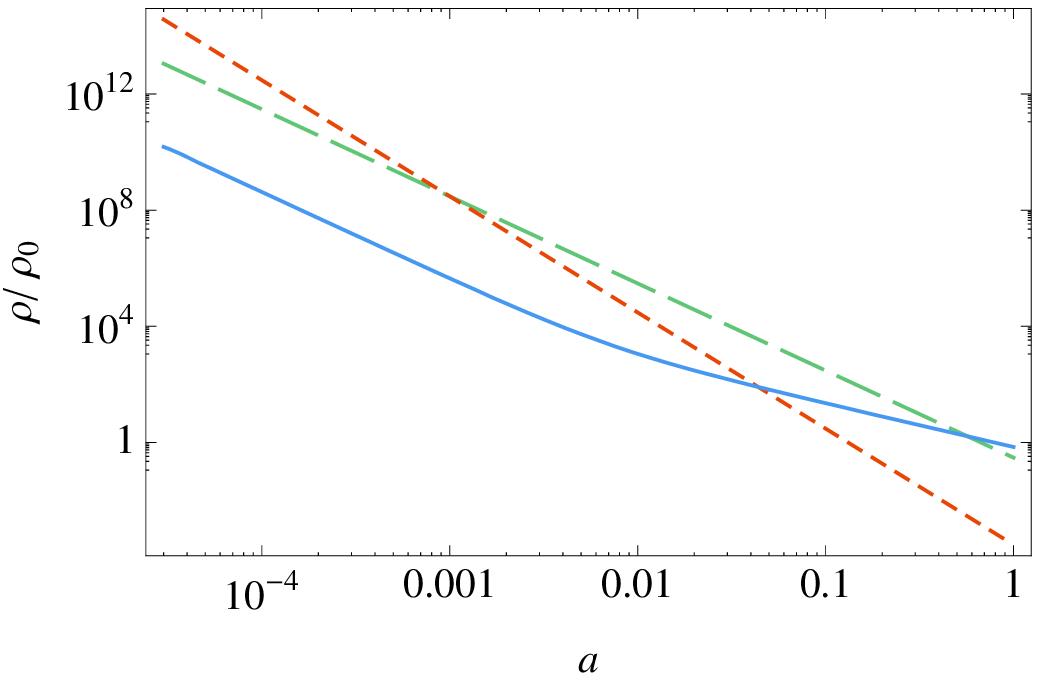}
\includegraphics[width=8.cm]{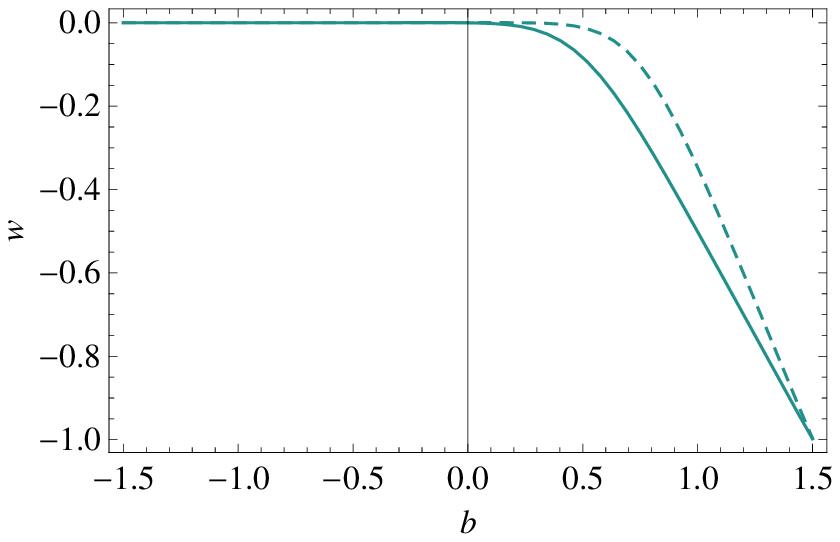}
\caption{In the left panel we show the cosmological evolution for the tower (solid blue line) where we can see that it tracks the matter behaviour in the early universe (we show radiation in short-dashed red and matter in long-dashed green ) and it exits that regime to eventually become the dominant component. In the right panel we plot the asymptotic equation of state of the tower as a function of $b$ in a matter (solid line) and a radiation (dashed line) dominated universes respectively. We can identify the range of values of $b$ for which we have the effective equation of state derived analytically in the main text in Eqs. (\ref{wlrad}) and (\ref{wlmat}).} 
\label{cosmevol}
\end{figure}

In the right panel of Fig. \ref{cosmevol} we show the  equation of state for the light modes (or, equivalently, the asymptotic equation of state parameter of the full tower) obtained by using the analytical solutions given in (\ref{Rsol}) and (\ref{Msol}) and, then, computing numerically the integral to obtain the corresponding energy density. In that figure we confirm the results found from our analytical estimates previously discussed. In particular, the linear dependence of $w_l$ on the spectral index given in (\ref{wlp}) that for a radiation dominated universe is
\be
w_l\vert_{p=1/2}=-\frac43 b+1\label{wlrad}
\ee
whereas, in a matter dominated universe, it reads
\be
w_l\vert_{p=2/3}=\frac12-b.\label{wlmat}
\ee
We also confirm the breakdown of these equations of state when they become larger than 0, i.e., for $b$ smaller than 1/2 and 3/4 in a matter and radiation dominated universes respectively. Thus, in order to have acceleration at late times $w_l<-2/3$ we need $b>7/6$. We shall see in the next section, that this condition is not really necessary to have accelerated solutions since, once the tower dominates, the effective equation of state of the tower will asymptotically approach -1.


From the previous discussion of the light and heavy modes evolution, we note that the total tower energy density can be approximately written as 
\be
\rho_\phi\simeq\rho_{ini}\left[\left(\frac{a_{ini}}{a}\right)^{3}+\left(\frac{H}{\sigma}\right)^{3-2b}\right]
\ee
where we have used (\ref{rhoh}) and (\ref{rhol1}). Since we are assuming that $H_{ini}\ll\sigma$, the dominant contribution at early times is given by the first term in the brackets, i.e., a matter-like behaviour. If we want the tower to dominate at late times, we need the second term to grow with respect to the first one. For a power law expansion $a\propto t^p$, this implies $b>\frac32 (1-p)$ which gives $b>\frac34$ for a radiation dominated universe and $b>\frac12$ for a matter dominated universe. On the other hand, by equalling both terms in the previous expression we can obtain the time $t_*$ at which the light modes contribution becomes dominant, which is given by
\be
\frac{t_*}{t_{ini}}\simeq\frac{H_{ini}}{H_*}\simeq\left(\frac{\sigma}{H_{ini}}\right)^{\frac{3-2b}{2b-3(1-p)}}
\ee
This analytical expression has also been confirmed numerically from the exact analytical solutions. In the case of a cosmological constant, dark energy becomes the dominant component of the universe when $H=\sqrt{2\Omega_\Lambda}H_0$. This gives us the approximate bound $H_*\gsim\sqrt{2\Omega_\Lambda}H_0$ for the light modes to dominate over the heavy ones early enough in the universe history.

\begin{figure}
\includegraphics[width=8cm]{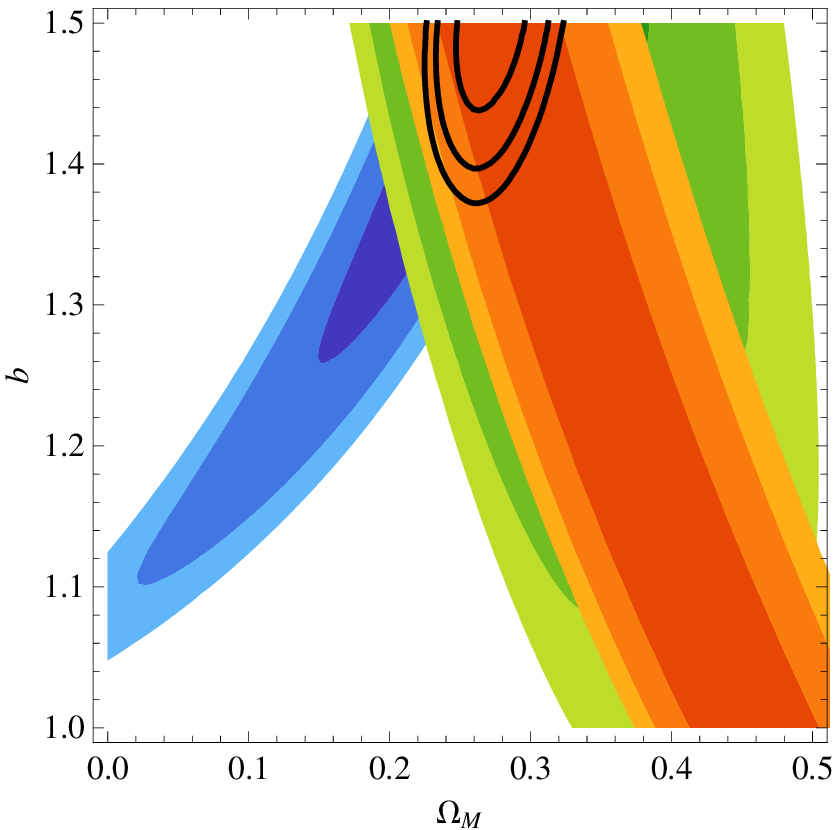}
\includegraphics[width=8cm]{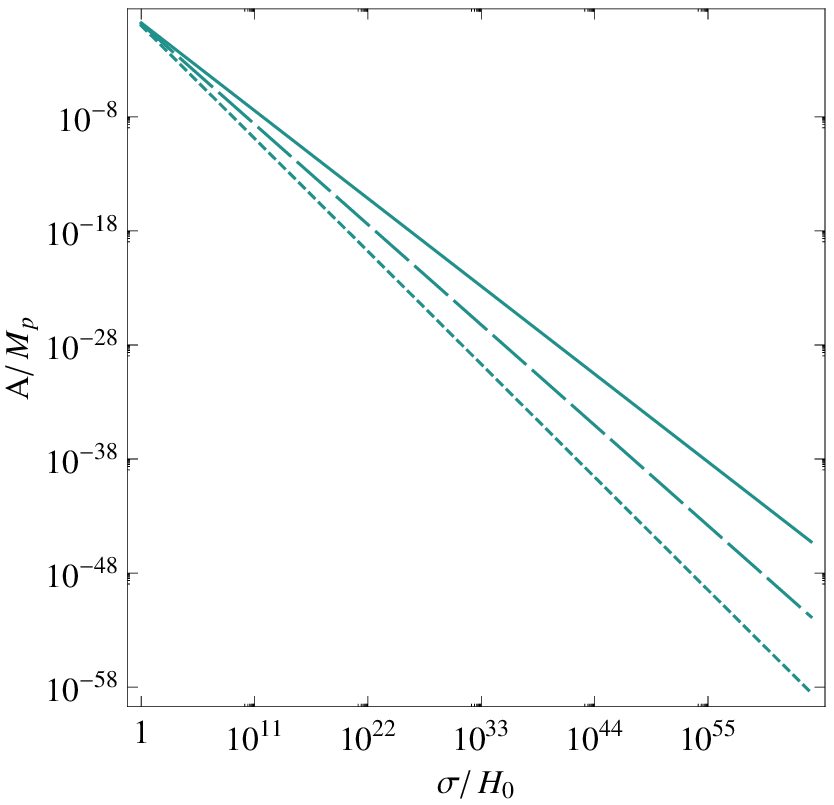}
\caption{The 68.27\%, 95\% and 99\% confidence regions obtained from the SNIa (blue), CMB distance priors (orange) from WMAP7 and BAO (green). The combined fit is shown by the solid black contours. The fit has been performed by assuming that the light modes behave like a perfect fluid with the equation of state given in (\ref{wlmat}) and that dark matter is some additional field, not the heavy modes of the tower. In the right panel we show the values of $A$ and $\sigma$ such that the light modes of the tower give the correct amount of dark energy today, given in Eq. (\ref{DEcondition}). We have used $\Omega_M=0.25$ and $b=1.2,1.3,1.4$ (solid, dashed and dotted respectively). The curves are not very sensitive to this value, being the major change due to $b$. }
\label{SNIa}
\end{figure}

If the light modes are to play the role of dark energy, then we need its relative abundance  to roughly accomplish the dark energy density parameter today, i.e, the following constraint must be satisfied 
\be
\Omega_l^0=\frac{8\pi G}{3H_0^2}\rho_l^0\simeq\frac{8\pi G A^2}{3\sqrt{2\pi}(3-2b)}\left(\frac{H_0}{\sigma}\right)^{1-2b}\simeq\Omega_{DE}\label{DEcondition}
\ee
This expression gives us the region in the parameter space in which the light modes of the tower can account for the current phase of accelerated expansion (see Fig. \ref{SNIa}). 

Since, the effective equation of state of the light modes  only depends on the parameter $b$ we can constrain its value from cosmological observations. We use the distance priors method as described in \cite{Komatsu:2008hk} to confront to CMB data \cite{Komatsu:2010fb}, the Union 2.1 SNIa compilation \cite{Suzuki:2011hu} and the 6 BAO measurements given by 6dF \cite{6dF}, SDSS \cite{SDSS} and WiggleZ \cite{WiggleZ}.  We show the corresponding confidence regions in Fig. \ref{SNIa}.
 To obtain such contours we have fitted the mentioned cosmological observations for a dark energy model with constant equation of state depending on $b$ as given in (\ref{wlmat}) and assuming spatially flat sections. We show the confidence regions in the $b-\Omega_M$ plane, which then determine the remaining parameters $A$ and $\sigma$ through the relation $\Omega_l^0=1-\Omega_M$.


\section{Tower dominated universe}
In the previous section we have shown how the tower can give a matter-like contribution from the heavy modes at early times and then the light modes can take over and give a dark energy contribution. It is interesting to note that the tower could play the role of both dark matter and dark energy. If the energy density associated with the heavy modes is initially (in the radiation dominated epoch) dominant over that of the light modes, then those modes will behave like dark matter. Notice that under the aforementioned assumptions,  a matter dominated universe is a consistent cosmological solution. This is so because, as we have seen in the previous section, the heavy modes behave like matter in a matter dominated universe. This is not very surprising, since it is well known that one massive scalar field oscillating around the minimum of the potential gives rise to a matter dominated universe. Thus, a collection of fields in such a regime, will also lead to a matter-like evolution. A difference between the two cases is that, in the case of one scalar field, the energy density oscillates and it is its average that redshifts as $(1+z)^3$. However, in the case of the tower, the averaging is effectively performed by the integration over the different masses so that no oscillations in the energy density evolution appear.

If we impose all the dark matter content of the universe be due to these heavy modes, then we need
\be
\Omega_h^0=\frac{8\pi G}{3H_0^2}\rho_h\simeq\frac{8\pi G \sigma^2A^2}{3H_0^2\sqrt{2\pi}(3-2b)(1+z_{ini})^3}\simeq\Omega_{DM}.\label{DMcondition}
\ee
The tower will give a matter behaviour until the energy density of the modes that were initially frozen are relevant and, eventually, become the dominant component. Unlike in the previous section, where the background evolution was assumed to be driven by some dominant barotropic fluid, in a tower dominated universe, when the light modes start dominating, its effective equation of state will tend towards -1.  This can however been understood from our findings in the previous section. While the heavy modes dominate, the light modes contribution will behave like a perfect fluid with equation of state parameter as given in (\ref{wlmat}). When the light modes of the tower start dominating the energy content of the universe, the Hubble expansion rate decays slower (since the dominant component, the light modes, has a smaller equation of state parameter) and this will make the lighter mass components start decaying later. This will make the energy density of the tower to decay slower and slower, so that the Hubble expansion rate will decrease slower and slower. This process will go on until both processes are in equilibrium, which will happen when the de Sitter evolution is reached.

\begin{figure}
\includegraphics[width=8cm]{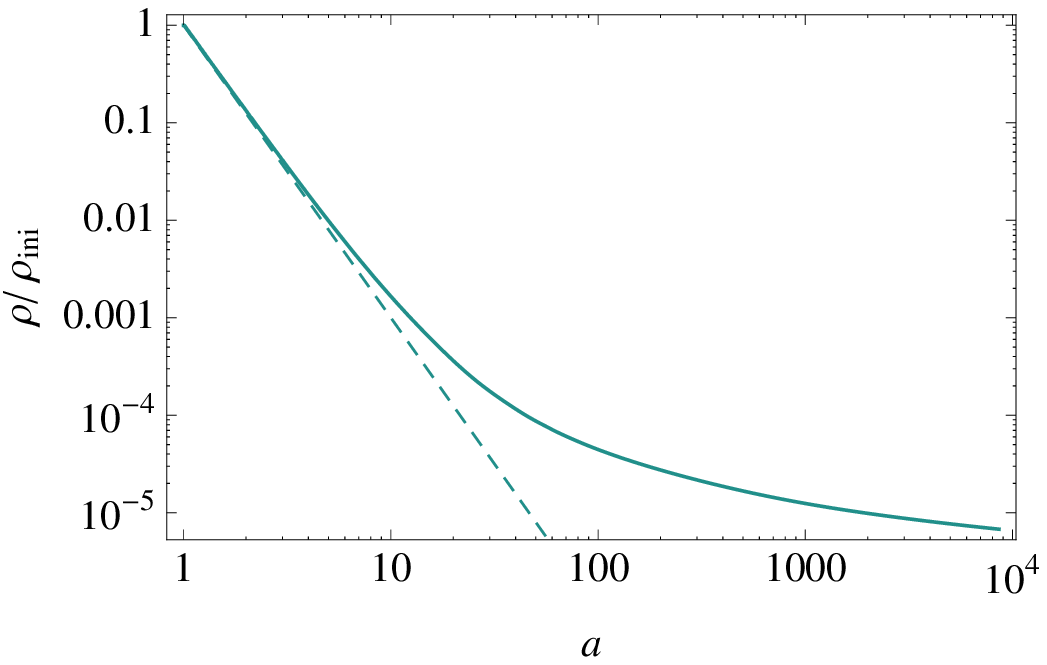}
\includegraphics[width=8cm]{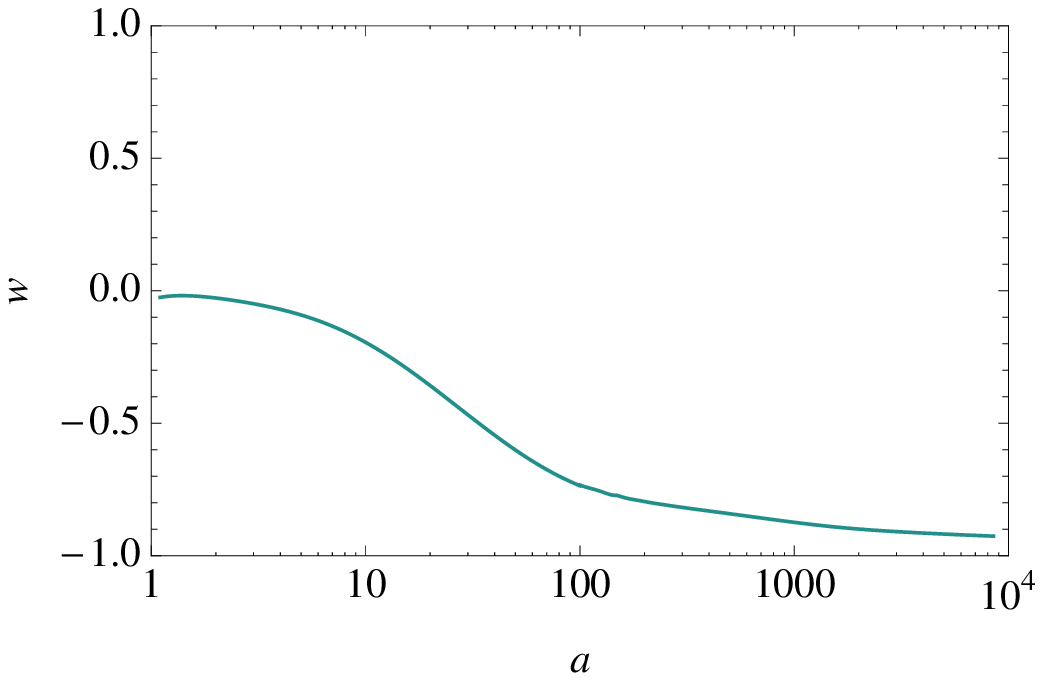}
\caption{In the left panel we plot  the tower energy density in a universe dominated by the tower. The parameters are such that the heavy modes dominate initially ($b=1$, $\sigma t_{ini}=10$), so we can see that the energy density decays like that of a matter fluid $\rho\propto a^{-3}$ (dashed line). At late times, the light modes take over and they asymptotically approach a (quasi) de Sitter regime. In the right panel we show the corresponding equation of state parameter where we see the initial matter-like regime ($w\simeq 0$) and the late time quasi de Sitter evolution with $w\simeq -1$.} 
\label{towerdom}
\end{figure}

Finding analytical solutions in the case when the Universe is dominated by the tower is much more involved than the case when the tower is subdominant. Thus, a numerical approach is required. To numerically solve the equations, we discretise the tower and solve the coupled system of Einstein and Klein-Gordon equations for the associated discrete tower. Of course, the final solution will not depend on the discretisation procedure, provided the partition is fine enough. We have checked that this is indeed the case for our case and, moreover, we can recover our analytical findings in the previous section when the tower is subdominant from our discretized associated problem. In Fig. \ref{towerdom} we show the obtained numerical solution for a universe dominated by the tower. We have chosen the initial conditions and model parameters so that the heavy modes are initially dominant and, thus, the tower behaves like a matter fluid. This is what we expected from our discussion above and has now been confirmed from our numerical solution. In this regime, the light modes behave like a perfect fluid with constant equation of state given by (\ref{wlmat}). Then, for $b>1/2$ we have that the effective $w$ for the light modes is smaller than that of the heavy modes and, therefore, their contribution will become dominant. At this point, the mechanism explained in the previous paragraph starts working and the de Sitter universe is eventually reached. We can see this in Fig. \ref{towerdom}, where the energy density scales like $a^{-3}$ initially and then it decays slower and slower, eventually becoming constant. In reality, the asymptotic behaviour is a quasi de Sitter expansion rather than a pure de Sitter. The same effect can be seen from the equation of state that is 0 initially and asymptotically approaches -1.

\section{Discussion}
In this work we have studied a cosmological model with a continuous tower of scalar fields. Rather than obtaining the model from a particular theory leading to a continuous tower, we adopt a phenomenological model-independent approach and have simply assumed a Universe with a continuous tower of massive scalar fields, whose mass spectrum is given by a Gaussian distribution peaking at the origin and with a certain dispersion $\sigma$. For the cosmological evolution of the tower,  we have used initial conditions according to a power law with amplitude $A$ and spectral index $b$. Then, the behaviour of the tower during the different epochs of the universe expansion history has been studied. We have found that the heavier modes of the tower give a matter-like contribution, as expected from the usual results of quintessence modes. However, for the light modes, we do not obtain the same behaviour as for one single light scalar field, that would give a $\Lambda$-like contribution today, provided its mass is $\lsim H_0$. Instead, we obtain that those modes cooperate to behave as a perfect fluid whose equation of state depends on the initial spectral index $b$ and ranges from 0 to $-1$. Only when $b\simeq\frac 32$, the light modes behave nearly like a cosmological constant.  We thus avoid having to introduce a very small mass like in the usual quintessence models in order to have acceleration today. The reason for this is that we have mass modes infinitely light in the tower that will remain frozen and, then, will give a constant energy density contribution, no matter how small the Hubble expansion rate is. In fact, this is the mechanism behind the effective equation of state parameter. As we mentioned in the introduction,  it must be noticed that, although the tower can give a nearly effective cosmological constant, the possible extra-dimensional origin of the tower (or any other context in which the tower could be generated) could induce an additional contribution to the cosmological constant. This is of course the usual cosmological constant problem that we do not intend to address in the present Letter.

We have obtained constraints on the parameters of the model from CMB, BAO and SNIa observations by modelling the light modes with a perfect fluid with constant equation of state. The tightest constraint is obtained for the spectral index $b$, while the initial amplitude and the distribution dispersion are degenerate. The reason is that these two quantities only contribute to the density parameter today of the light modes, while the effective equation of state is entirely determined by $b$. 

Finally, we have studied the case of a universe dominated by the tower. In the case when the heavy modes dominate, the universe behaves as dominated by a matter-like component, as expected. However, for appropriate values of the spectral index $b$, the light modes will eventually dominate and their effective equation of state approaches -1 so that the universe eventually becomes de Sitter.

We can also comment on the fact that having a continuous tower with  all possible masses, including a massless component might be troublesome. If a massless (or very light) scalar field is coupled to standard model particles, then it will mediate a long range force that has never been detected in fifth force experiments. Ways out to this are usually given by making the field dynamics environment dependent so that its effective potential (or mass) screens the field effects in appropriate environments (chameleon or symmetron mechanism) or non-linear terms become important in high density regions (Vainhstein mechanism). These mechanisms could be implemented for the individual modes of our tower to avoid fifth force constraints. Another potential way of avoiding such tests could be developed by taking advantage of the distribution weighting the different mass modes. Thus, one could try to make the mass distribution environment dependent such that it strongly suppresses the lightest modes in regions of high density and, then, no fifth forces will arise in local gravity tests from the tower.


\section*{Acknowledgments}

J.B.J. is supported by the Wallonia-Brussels Federation grant ARC No.~11/15-040 and also thanks support from the Spanish MICINNÕs Consolider-Ingenio 2010 Programme under grant MultiDark CSD2009-00064 and project number FIS2011-23000. 
D.F.M. thanks the Research Council of Norway
FRINAT grant 197251/V30. D.F.M. is also partially
supported by project $CERN/FP/123618/2011$ and $CERN/FP/123615/2011$.  The work of P.S. is sponsored by FCT - Funda\c{c}\~ao para a Ci\^encia e Tecnologia, under the grant SFRH / BD / 62075 / 2009. P.S.  is very thankful to the Departamento de F\'isica Te\'orica I of the Universidad Complutense de Madrid for their kind hospitality.


\end{document}